\documentclass[useAMS,usenatbib]{mn2e}
\usepackage{amsmath}
\usepackage{amsbsy}
\usepackage{times}
\usepackage{graphicx}
\usepackage{aas_symbols}
\usepackage{threeparttable}
\usepackage{epstopdf}
\usepackage{url}
\usepackage{cases}
\usepackage{color,ulem}
\usepackage{multirow}

\begin{document}
	
	\title[Merger time of binary neutron stars]{The Gravitational waves merger time distribution of binary neutron star systems}
	
\author[P. Beniamini $\&$ T. Piran]{
Paz Beniamini$^{1,2}$\thanks{E-mail: paz.beniamini@gmail.com}
$\&$ Tsvi Piran$^{3}$
\\
% List of institutions
$^{1}$Department of Physics, The George Washington University, Washington, DC 20052, USA \\
$^2$Astronomy, Physics and Statistics Institute of Sciences (APSIS)\\	
$^3$Racah Institute of Physics, the Hebrew University, 91904, Jerusalem, Israel}
	
%\date{Accepted; Received; in original form ...}
	
%\pubyear{2019}
	
\maketitle
	
\begin{abstract}
	Binary neutron stars (BNS) mergers are prime sites for $r$-process nucleosynthesis. Their rate determines the chemical evolution of heavy elements in the Milky Way. 
	The merger rate of BNS is a convolution of their birth rate and the gravitational radiation spiral-in delay time. Using the observed population of Galactic BNS we show here that the lifetimes of pulsars in observed BNSs are sufficiently short that the ages of BNSs have little to no effect on the observed merger time distribution. We find that at late times ($t\gtrsim 1$ Gyr) the gravitational wave delay time distribution (DTD) follows the expected $ t^{-1}$. However, a significant excess of rapidly merging systems (between $40-60\%$ of the entire population) is apparent at shorter times.
	Although the exact shape of the DTD cannot be  determined with the existing data, in all models that adequately describe the data we find at least $40\%$ of BNSs with merger times less than 1Gyr.
	This population of rapid mergers implies a declining deposition rate of $r$-process materials that is consistent with several independent observations of heavy element abundances in the Milky Way.
	At the same time this population that requires initial binary separations of roughly one solar radius clearly indicates that these binaries had common envelope progenitors. Our results suggest that a significant fraction of future LIGO/Virgo BNS mergers would reside in star forming galaxies.  
\end{abstract}
%%%%%%%%%%%%%%%%%%%%%%%%%%%%%%%%%%%%   
\begin{keywords}
	gravitational waves -- stars: neutron -- Galaxy: abundances -- stars: evolution 
\end{keywords}
%%%%%%%%%%%%%%%%% BODY OF PAPER %%%%%%%%%%%%%%%%%%
\section{Introduction}
\label{sec:Intro}
The merger of binary neutron star (BNS) systems results in a blast of gravitational wave (GW) radiation, in a short gamma-ray burst (GRB) and in the synthesis of $r$-process elements at a rate that can account for the overall formation of those elements in the Universe. The connection between these different aspects of BNS mergers  \citep{Lattimer1974,Lattimer1976,Eichler1989} has recently been given very strong support due to the detection of GWs from GW170817 which were accompanied  by a GRB \citep[albeit an unusually weak one, see e.g.][]{Kasliwal2017,Matsumoto2019}, a macronova (powered by the radioactive decay of $r$-process elements) and a multiwavelength afterglow. 
Interestingly the implied rate and yield of heavy $r$-process element is consistent with that required for the production of heavy $r$-process elements in the Galaxy \citep[see e.g.][for a recent review]{HBP2018}.

The rate of BNS mergers depends on the birth rate of these systems as well as on the delay times from birth to merger, commonly known as the `delay time distribution' (DTD). As neutron stars' progenitors are massive stars their birth rate follows the star formation rate (SFR) with a minimal delay. The DTD is dominated by this GW spiral-in time. As a BNS loses energy due to GW radiation its orbit shrinks and eventually the neutron stars merge. The time until merger $t_{\rm m}\propto a^4 (1-e^2)^{7/2}$ depends on the initial semi major axis, $a$, and the eccentricity, $e$, of the BNS. As such the determination of the DTD and in particular of the minimal time delay, $t_{\rm min}$ will provide valuable information on the unknown final stages of evolution of BNS progenitors. 
Here we explore the DTD  using the observed Galactic population of BNS systems. This approach is complementary to attempts to explore the DTD by comparing the rate of short GRBs (believed to be associated with BNS mergers) and the SFR \citep[see e.g.][]{Guetta2005,Guetta2006,Guetta2009,Leibler2010,Dietz2011,Coward2012,WP2015}. 
An independent single observation arises from GW170817. The host galaxy of this event is an S0-type galaxy with a very low star formation rate, suggesting a time delay between the BNS formation and merger of $\approx 1-10$ Gyr.

The GW merger time, $t_{\rm m}$, depends strongly on $a$, and since many BNSs exhibit low orbital eccentricities, it is insightful to consider the expected DTD resulting from an initial power-law distribution of semi major axes, $dN/da\propto a^{-n}$ with circular orbits ($e=0$). The result is $dN/dt_{\rm m}\propto t_{\rm m}^{-(n+3)/4}$. In particular, if the semi-major axis is uniformly distributed in log-space, $dN/d\log a=const$, one gets the well known result $dN/dt_{\rm m}\propto t_{\rm m}^{-1}$ \citep[see e.g.][]{Piran1992,Yungelson2000,Guetta2005,Guetta2006,Totani2008,Maoz2014}. Note also that due to the strong dependence of $t_{\rm m}$ on $a$, the distribution of semi-major axes must be very steep in order to get significantly steeper $dN/dt_{\rm m}$ profiles.

We use the observed distribution of merger times of the Galactic BNSs to derive the DTD in a model-independent way, directly from observations. The paper is structured as follows. In \S \ref{sec:GalacticBNS} we describe the observed sample. We then study analytically the connection between the distribution of delay times at birth and the merger time (and age) distribution viewed at some later time in \S \ref{sec:direct}. In \S \ref{sec:like} we perform a likelihood analysis in order to determine the underlying DTD from the observed population. We examine some implications on the total number of un-merged systems and observable pulsars remaining in BNSs in the Galaxy today in \S \ref{sec:numbers}. We compare our results with other findings in \S \ref{sec:other} and we summarize and discuss some broader implications in \S \ref{sec:discussion}.

\section{Galactic BNS observations}
\label{sec:GalacticBNS}
Our sample includes the 15 observed field Galactic BNS systems. We exclude two Galactic BNSs that reside in globular clusters. Table \ref{tbl:sample} provides the semi-major axes, eccentricities, merger times and spin-down times of these systems. The observed merger time distribution is shown in figure \ref{fig:tageeffect}. 

Clearly this  sample  does not include all Galactic BNSs, as the observed systems are only those that have an active pulsar pointing towards us. Since pulsars live a finite time there are many systems that cannot be observed. The BNS pulsars are mostly recycled ones \citep{BP2016,Tauris2017} and their true ages\footnote{We consider as the true age of the system the time since the formation of the second neutron star.} may be considerably younger than  their characteristic dipole spin-down times, $\tau=P/2\dot P$, \citep{Giguere2006,Kiziltan2010,Oslowski2011}. 
This is because for any pulsar with a braking index $n>1$, the ratio $\tau=P/(n-1)\dot P\propto P^{n-1}$ increases with $P$, and always remain an upper limit on the true age \citep[see e.g][for  more details]{ShapiroTeukolsky1986,Beniamini2019}.
For example, in the double pulsar system, J0737-3039, the two pulsars (which are expected to be of a similar age) have spin-down ages of 210 and 50 Myr \citep{Kramer2006}. This is much longer than what is expected to be their true age difference. 

\begin{table}
	%	\begin{center}
	\caption{Parameters of Galactic field binary neutron stars.}
	\resizebox{0.4848\textwidth}{!}{
		\begin{threeparttable}	\begin{tabular}{cccccc}\hline	
				System &  $a [10^{11}\mbox{ cm}]$  & $e$ & $t_{\rm m} [\mbox{ Gyr}]$ & $P/2\dot P[\mbox{ Gyr}]$ & ref. \\ \hline
				J1946+2052 & 0.73 & 0.064 & 0.05 & 0.3 & 16 \\
				J1757-1854 & 1.32 & 0.605 & 0.07 & 0.13 & 14 \\
				J0737-3039 & 0.88 & 0.088 & 0.086 & 0.05/0.21 &1\\
				J1906+0746 & 1.22 & 0.085 & 0.31 & $10^{-4}$ &2,3\\
				B1913+16 & 1.95 & 0.617 & 0.38 & 0.08 & 6\\
				J1913+1102 & 1.45 & 0.089 & 0.47 & 2.7 & 13 \\
				J0509+3801 & 2.16 & 0.59 & 0.69 & 0.15 & 17 \\
				J1756-2251 & 1.87 & 0.181 & 1.61 & 0.44 & 4,5\\
				J1829+2456 & 4.48 & 0.139 & 55.4 & 12.4 & 8 \\
				J1411-2551 & 7.58 & 0.169 & 460 & 10.4 & 15 \\
				J0453+1559 & 10.4 & 0.112 & 1400 & 2.6 & 11 \\
				J1811-1736 & 28.3 & 0.828 & 6400 & 1.8 & 10 \\
				J1518+4904 & 17.2 & 0.249 & 8900 & 10.4 & 9 \\
				J1930-1852 & 50.9 & 0.4 & $5\!\times\! 10^5$ & 0.16 & 12 \\
				\hline  
				\label{tbl:sample}
			\end{tabular}
			\begin{tablenotes}
				\item [(a)] References (ordered by table numbers): \cite{Kramer2006,Lorimer2006,vanLeeuwen2015,Faulkner2005,Ferdman2014,Weisberg2010,Stairs2002,Champion2005,Janssen2008,Corongiu2007,Swiggum2015,Martinez2015,Lazarus2016,Martinez2017,Stovall2018} 
			\end{tablenotes}
		\end{threeparttable}
		%		\end{center}
	}
\end{table} 

\section{General Considerations}
\label{sec:direct}
\subsection{The delay time distribution - DTD}
\label{sec:DTD}
The merger time distribution, $dN/dt$, is a convolution of the intrinsic DTD, $D(t)$, with the BNS birth rate, $r(t)$, where time here is measured backwards from the present day (i.e. today we have $t=0$ and at earlier times, $t$ is larger): 
\begin{equation}
\label{eq:dNdt0}
\frac{dN}{dt}=\int_0^{t_{\rm H}} r(t')D(t'+t)dt', 
\end{equation}
where $t_{\rm H }$ is the Hubble time. We will consider DTDs that have a minimal $t_{\rm min} $ and maximal $t_{\rm max}$ range. 

The distribution $dN/dt$ is different from the observed merger time distribution  $dN_{\rm obs}/dt$ since, as mentioned earlier, the observed distribution depends on detection of the corresponding pulsars and those have a finite lifetime. 
To account for the finite lifetimes of the pulsars in the BNS systems we introduce a typical lifetime $t_{\rm a}$ in which one of the pulsars in the BNS system can still be observed.  In reality, there is a distribution of lifetimes. As there is not enough information today to reveal this distribution we will take $t_{\rm a}$ to  be a constant. If there is a distribution of lifetimes than $t_{\rm a,min} < t_{\rm a} < t_{\rm a,max}$ the transition between the regimes discussed below will be gradual, but the basic qualitative features will remain. The regime $t\ll t_{\rm a}$ ($t\gg t_{\rm a}$) should then be interpreted as $t\ll t_{\rm a,min}$ ($t\gg t_{\rm a,max}$). We show later (see \ref{sec:Ages}) that the value of $t_{\rm a}$ is likely small compared to the typical delay times. However, for the sake of generality, we make no assumptions on the value of $t_{\rm a}$ in the following discussion.

\begin{figure}
	\centering
	\includegraphics[width=0.39\textwidth]{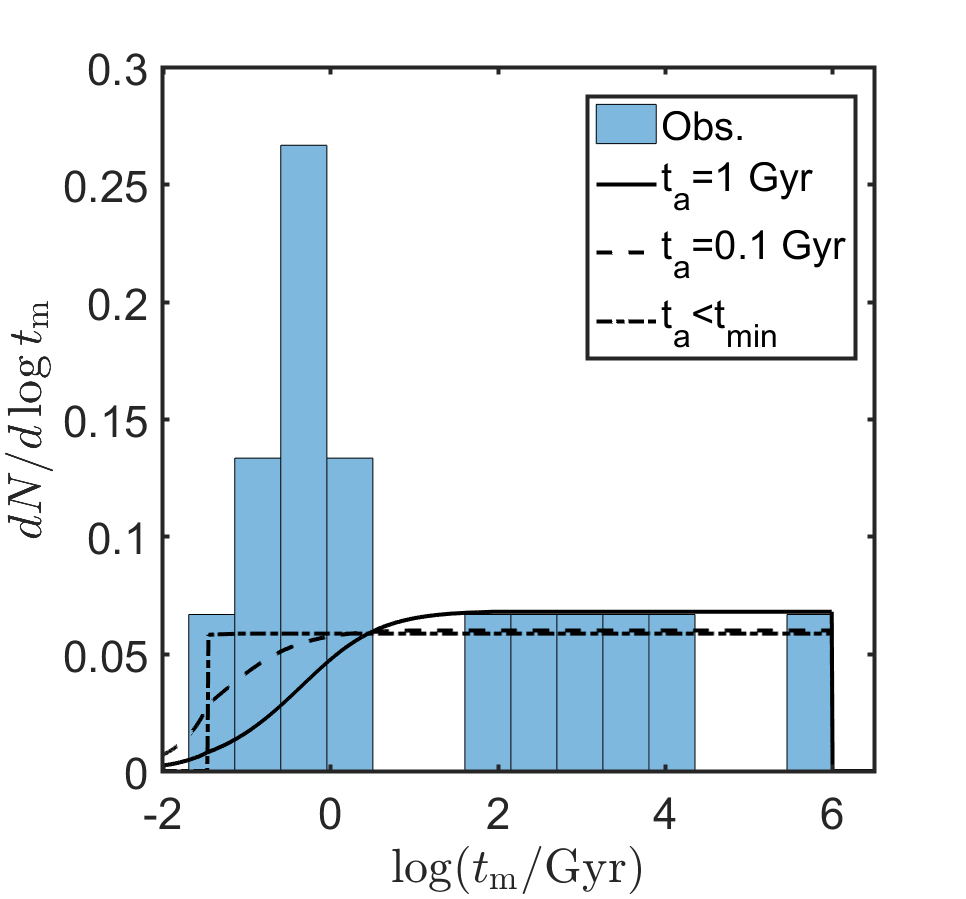}
	\caption{Merger time distributions of observed BNSs as compared with a delay time distribution $D(t)\propto t^{-1}$ between $t_{\rm min}=0.035$ Gyr and $t_{\rm max}=10^6$ Gyr with different values of the maximum pulsar age, $t_{\rm a}$. A statistically significant excess of systems with merger times $\lesssim 1$\,Gyr is required by observations, whereas the finite age effect can only {\it decrease} the number of short merger time systems.}
	\label{fig:tageeffect}
\end{figure}

We write the number of observed BNSs with a given time to merger, $dN_{\rm obs}/dt$, as:
\begin{equation}
\label{eq:dNdt0}
\frac{dN_{\rm obs}}{dt}=\int_0^{t_{\rm a}}r(t')D(t'+t)dt'.
\end{equation}
Since, as we show later, the life times of the observed BNSs as active pulsars is much shorter than the scale on which the Galactic star formation rate (SFR) changes  we use here a constant Galactic SFR, $r(t)=r_0$. This simple assumption has been suggested in the literature as a good approximation for the star formation rate of the Milky Way (MW) during the majority of its lifetime \citep[see e.g.][]{Snaith2015}.  For $r(t)=r_0$  and for $t_{\rm a}<t_{\rm min}$, the observed merger distribution is:
\begin{eqnarray}
\label{eq:tage}
\frac{dN_{\rm obs}}{dt}\propto \! \left\{ \begin{array}{ll}\! 0 & \mbox{ for } t<t_{\rm min}-t_{\rm a}\\ \! const & \mbox{ for } t_{\rm min}-t_{\rm a}<t\ll t_{\rm min}\\
\! D(t)  & \! \! \mbox{ for }t\gg t_{\rm min}.
\end{array} \right.
\end{eqnarray}
Note that for $t_{\rm a} \ll t_{\rm min} < t $ the observed distribution is simply proportional to the DTD:  $dN_{\rm obs}/dt\propto D(t)$ regardless of the functional form of $D(t)$. Consider now the opposite limit, $ t_{\rm a}>t_{\rm min}$ (assuming once more a constant formation rate). For $t> t_{\rm a}$, that is for 
merger times much longer than the age of the systems, the pulsars stop shining on a time scale much shorter than their merger time. Hence, $dN_{\rm obs}/dt \propto D(t)$ as before. 
But, of course, there are many more ($t_{\rm H}/t_{\rm a}$) BNS systems that we cannot observe. Alternatively, at much shorter times, $t\ll t_{\rm a}$ the observed population is reduced compared to an extrapolation from the $t\gg t_{\rm a}$ regime since only a fraction $t/t_{\rm a}$ of the systems born in the last $t_{\rm a}$ years with a delay time $t$ have not merged yet. $dN_{\rm obs}/dt$ is therefore proportional to $(t/t_{\rm a})D(t)$. At delay times even lower than $t_{\rm min}$, the only systems observable are the tail of the systems born with a delay of $\sim t_{\rm min}$ (which by construction are the type of systems that dominate the intrinsic population) that were formed $t_{\rm min}-t$ years ago. For $t\ll t_{\rm min}$ the two latter factors are constant, and thus $dN_{\rm obs}/dt$ is constant.

Consider, for example, a power-law DTD:
\begin{eqnarray}
D(t)\!=\frac{1\!-\!b}{t_{\rm max}^{1-b}-t_{\rm min}^{1-b}} ~ t^{-b} \quad {\rm for } ~~t_{\rm min}\le t \le t_{\rm max} . 
\end{eqnarray}
For  an index $b> 1$ and $ t_{\rm a}>t_{\rm min}$ we obtain
\begin{eqnarray}
\label{eq:dNdt}
\frac{dN_{\rm obs}}{dt}\propto 	\left\{ \begin{array}{ll}const & \! \! t\ll t_{\rm min}\\
t^{1-b} \! \! & \! \! t_{\rm min} \ll t\ll t_{\rm a} \\ t^{-b} \! \! & \! \!  t_{\rm a}\ll t\ll t_{\rm max}.
\end{array} \right.  
%\nonumber
\end{eqnarray}
Overall we find that the finite lifetime of pulsars implies that for merger durations longer than $t_{\rm a}$ the observed merger time distribution simply follow the DTD. For durations shorter than $t_{\rm a}$ we observe fewer systems than what the DTD would predict.

\subsection{The Galactic BNS distribution} 
The observed distribution of Galactic BNS merger times is shown in figure \ref{fig:tageeffect}.
At times much larger than the typical ages of the pulsar, $dN_{\rm obs}/dt \propto D(t)$, i.e. it is directly related to the DTD.
It is evident from figure \ref{fig:tageeffect} that at $t\gtrsim 1$\,Gyr, $dN/d\log t_{\rm m}$ is approximately constant, suggesting $D(t) \propto t^{-1}$. However, at lower merger times we find an observed excess of systems with $0.05 \mbox{ Gyr}\lesssim t_{\rm m}\lesssim 1$\,Gyr (seven out of fifteen systems, representing a fraction $Z_{<1 Gyr}= 0.47$ of the population), as compared with the $D(t)\propto t^{-1}$ model predictions (for which  $Z_{<1 Gyr}=0.19$, as we discuss later in \S \ref{sec:justPL}). At the same time, as we argued above, the finite lifetime of the pulsars leads instead to a suppression of the number of systems with $t_{\rm m} < t_{\rm a}$.

$D(t)\propto t^{-1}$ is a good description of the observations for $t>1$\,Gyr. The excess of short duration mergers (relative to predictions of the $D(t)\propto t^{-1}$ model) at $t_{\rm m}\lesssim 1$\,Gyr must be an intrinsic feature of the DTD. 
Note that the possibility  that this excess is due to an increase of the BNS formation rate in the last $1$\,Gyr is highly unlikely. This time-scale  is much smaller than the scale on which significant Galactic changes took place. 

\begin{figure}
	\centering
	\includegraphics[width=0.39\textwidth]{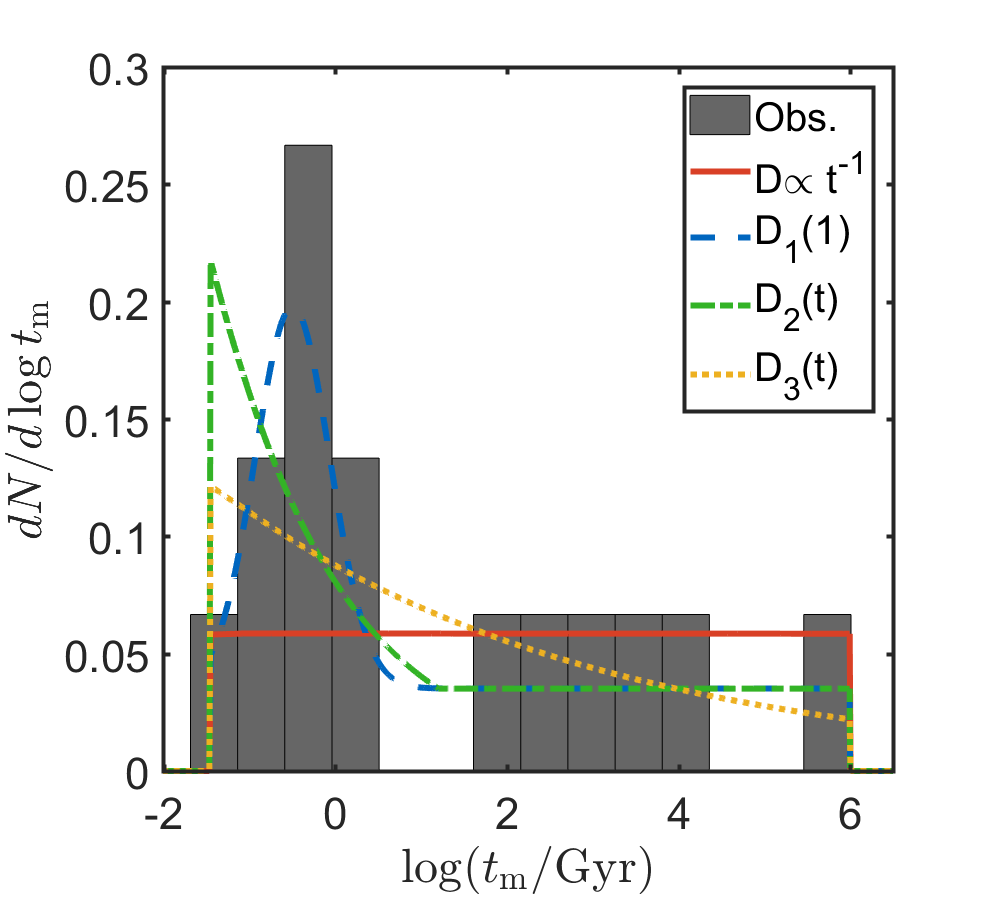}\\
	\includegraphics[width=0.39\textwidth]{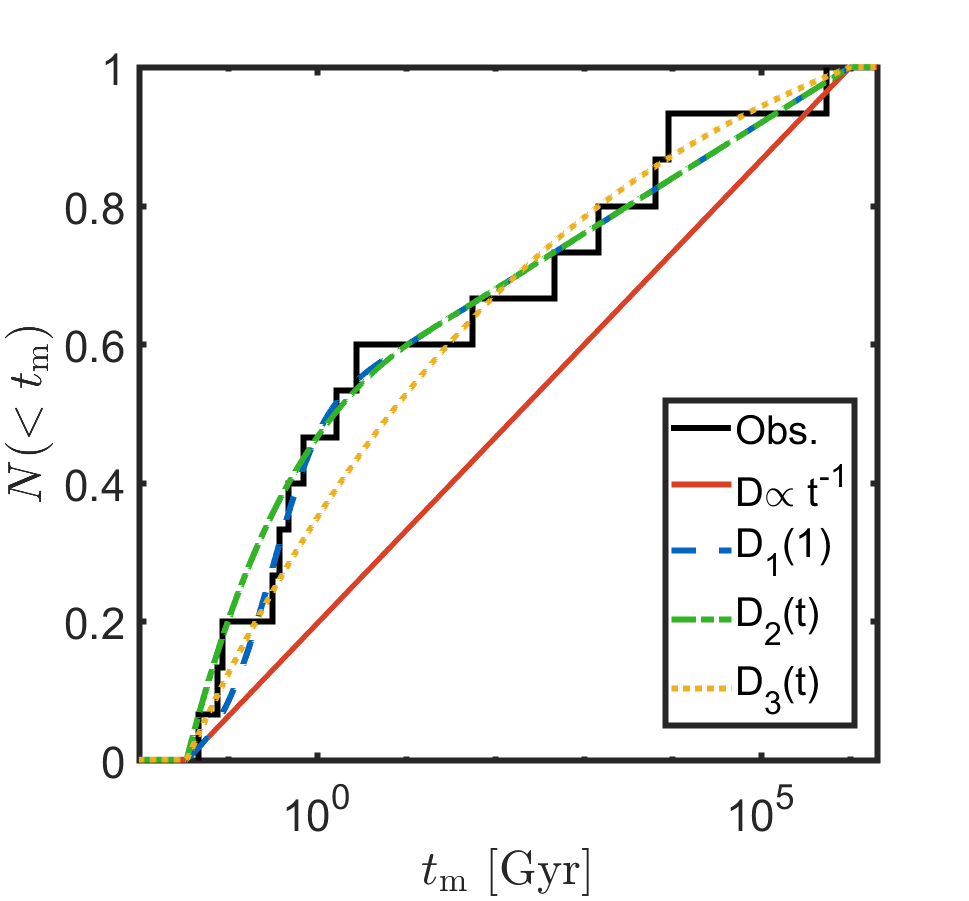}
	\caption{Merger time distributions of observed BNSs as compared with the four different delay time distribution models described in \S \ref{sec:tm1}, \ref{sec:justPL}, \ref{sec:shortcomp}. In all cases we take $t_{\rm a}<t_{\rm min}$ and DTD model parameters as described in \S \ref{sec:like}. A statistically significant excess of systems with merger times $\lesssim 1$\,Gyr is required by observations.
		Top: the differential distribution, Bottom:  the integrated distribution.}
	\label{fig:modelcompare}
\end{figure}

\subsection{Ages of observed BNS systems}
\label{sec:Ages}
As mentioned in \S \ref{sec:GalacticBNS}, the ages of BNS systems are less or equal than their spin-down times. The distribution of observed ages of BNS systems, is naturally affected by the DTD (at older ages, a larger fraction of systems have already merged). The number of systems with age $t_{\rm age}$ is given by
\begin{eqnarray}
\label{eq:tage}
\frac{dN}{dt_{\rm age}}\!&=&\!  r(t_{\rm age})\!\int_{t_{\rm age}}^\infty D(t) dt \\ \nonumber 
&\propto& \! \left\{ \begin{array}{ll}\! const & \mbox{ for } t_{\rm age}<t_{\rm min}\\
\! t_{\rm age}^{1-b}  & \! \! \mbox{ for }t_{\rm min} < t_{\rm age} < t_{\rm max}.
\end{array} \right.
\end{eqnarray}
where  we have assumed in the second line a constant formation rate and a power-law DTD with $b>1$.

One of the binary systems, J1906+0746, has a very short spin down time, $P/2\dot P=10^{-4}$\,Gyr. 
This value is much smaller than the smallest  observed delay time (see table \ref{tbl:sample}).
If indeed $P/2\dot P$ reflects  an upper limit on the real age of this pulsar then, independently of the shape of the delay time distribution,  
equation \ref{eq:tage} implies that there should be many older systems.
As an illustration, there should be roughly 100 systems with $P/2\dot P<0.01$\,Gyr for each system with $P/2\dot P<10^{-4}$\,Gyr. Even taking into account a possible observational bias according to which older system may be harder to detect this ratio is inconsistent with the observations. 

To explore the implication of this inconsistency we consider the  $P/2\dot P$ values in table \ref{tbl:sample} as upper limits on the true ages of the systems. We assume that the systems live for a fixed time $t_{\rm a}$  and search using a Monte Carlo simulation for a value of $t_{\rm a}$ for which a simulated set of $N$ ages (where $N$ is the number of observed BNS systems) resulting from a given $b$,  $t_{min}$ and $t_a$ is consistent with these upper limits. Namely we look for the maximal value of $t_a$ for which more than 5\% of the trials are consistent. Larger values of $t_a$ are ruled out at $2 \sigma$.

For $0.01 \mbox{ Gyr}\leq t_{\rm min} \leq 0.05\mbox{ Gyr}$ and $1\leq b \leq 1.5$, as suggested by the observed merger times \footnote{The results hold as long as there is no significant drop in the age distribution given by Eq. \ref{eq:tage} above $t_{\rm min}$.}, we find $t_{\rm a}<0.03$\,Gyr at a $3\sigma$ confidence level. This result, which depends weakly on the choice of $t_{\rm min}$ and $b$ depends very strongly on the very short $P/2\dot P$ of  J1906+0746. 
However, even if the spin-down age of the pulsar in J1906+0746 is disregarded as an anomaly, the remaining distribution of spin-down ages still constrains $t_{\rm a}<0.3\mbox{ Gyr}$ at a 2$\sigma$ level. That is even if $t_{\rm a}>t_{\rm min}$,  the dynamical range between the two cannot be large. Therefore, the effect of $t_{\rm a}$ on the observed distribution must be relatively small. This conclusion is  consistent with the estimate provided in \S \ref{sec:numbers} based on the number of observed Galactic BNS systems in comparison with estimates of the BNS formation rate.
As discussed in \S \ref{sec:DTD}, this means that the observed distribution of merger times roughly traces the observed ones, and that the exact value of $t_{\rm a}$ is not required to deduce the shape of the DTD.

Following these results we assume in the following that $t_{\rm a}<t_{\rm m}$. This means that the orbits of the majority of systems don't evolve appreciably during the time  that their pulsars remain observable and hence the observed merger time distribution reflects the intrinsic one. We also verify later (see \ref{sec:lifetimeLike}), using a maximum likelihood analysis, this assumption. 

\section{Likelihood analysis}
\label{sec:like}
We consider several models that can be fitted to the observed merger time distribution. Our models involve one or two parameters (denoted here as $X_i$ for the more general case) that describe the DTD.
We search for the values of $X_i$ that maximize the likelihood of obtaining the observed distribution of Galactic BNS merger times.
To do so we construct a likelihood function as follows
\begin{equation}
\label{eq:Likephen}
%&L(t_{\rm a},X_i)=\Pi_j P(t_{\rm age}|t_{\rm a}) P(t_{\rm merg,j}+t_{\rm age}|X_i)
L(X_i)=\Pi_j  P(t_{\rm m,j}|X_i)
\end{equation}
where $j$ goes over the observed systems in the sample, $P(t_{\rm m,j}|X_i)$ is the probability that a system was born with a delay time $t_{\rm m,j}$ given the delay time distribution specified by $X_i$. 

\subsection{A $t^{-1}$  delay time distribution}
\label{sec:tm1}
Our first choice for the DTD is a simple power-law $t^{-1}$. This is motivated, 
as mentioned in \S \ref{sec:Intro}, by the strong dependence of $t_{\rm m}$ on the initial BNS separation  \citep[see e.g.][]{Piran1992,Yungelson2000,Guetta2005,Guetta2006,Totani2008,Maoz2014}. Moreover, an inspection by eye of the long merger time part of the distribution (which is not influenced by pulsar's lifetimes and the choice of the edges of the distribution) suggests that this model is consistent at least for long merger durations.
The observed delay times depend only weakly  on $t_{\rm max}$ which we fix here to be $t_{\rm max}=10^6$\,Gyr. 
The DTD is in this case a function of a single parameter $t_{\rm min}$. The likelihood  is maximized for $t_{\rm min}=0.035$\,Gyr. 
Using a Kolmogorov-Smirnov (KS) test, we compare the observed merger time distributions with the model. The two distributions are shown in figure \ref{fig:modelcompare}. The KS test rules out the $t^{-1}$ model at a 2$\sigma$ confidence level. As mentioned in \S \ref{sec:direct}, there is a much larger fraction of short mergers in observations ($Z_{<1 \rm Gyr}=0.47$) than expected for the best $D(t)\propto t^{-1}$ ($Z_{<1 \rm Gyr}=0.19$). This is the main source of discrepancy between this model and observations. It is important to stress that, as we have seen in \S \ref{sec:DTD}, introducing longer lifetimes will only exacerbate the problem as this leads to a decrease in the number of observed systems with short merger times (see figure \ref{fig:tageeffect}).
Given this discrepancy we explore in the following several DTD models that involve a second free parameter.

\subsection{A general power-law delay time distribution}
\label{sec:justPL}
The observed merger times clearly span many orders of magnitude (see table \ref{tbl:sample}). The simplest  functional form spanning a wide range is  a power-law.
A natural generalization is to use another power-law $t^{-b}$ and let $b\neq 1$ be a free parameter in addition to $t_{\rm min}$. Due to the vast range between $t_{\rm min}$ and $t_{\rm max}$ (at least seven orders of magnitude; see table \ref{tbl:sample}) even a small change in $b$ can have a significant effect on the observed distribution. Indeed taking $b=1.1, t_{\rm min}=0.035 \mbox{ Gyr}$ (and  $t_{\rm a}<t_{\rm min}$) results in a distribution (denoted in figure \ref{fig:modelcompare} as $D_3(t)$) that is consistent with  the current data. The main importance of this possibility is the simplicity of implementing it in analytic expressions such as in \S \ref{sec:direct}, \ref{sec:numbers}.

\subsection{A short delay time population}
\label{sec:shortcomp}
At long merger times $\gtrsim \!1$\,Gyr, the DTD is well described by $D(t)\propto t^{-1}$. Motivated by the excess of events with short merger times,  we add a short merger time log-normal component to the $t^{-1}$ power-law: 
\begin{equation}
D_1(t)=At^{-1} + B{\ \rm  Lognormal}(\mu,1) \mbox{ for }t_{\rm min}<t<t_{\rm max}\ .
\end{equation}
The second term is a log-normal distribution with $\sigma=1$ (implying a width that is of the same order of magnitude as the median). Based on the observed sample and on the modeling for the $t^{-1}$ and general power-law DTDs we fix here $t_{\rm min}=0.035$ Gyr, $t_{\rm max}=10^6$ Gyr. This model involves two free parameters: $\mu, f$, where $f$ is the ratio of events resulting from the log-normal distribution  out of the total number of events (and is proportional to $B/(A+B)$). 

\begin{figure}
	\centering
	\includegraphics[width=0.39\textwidth]{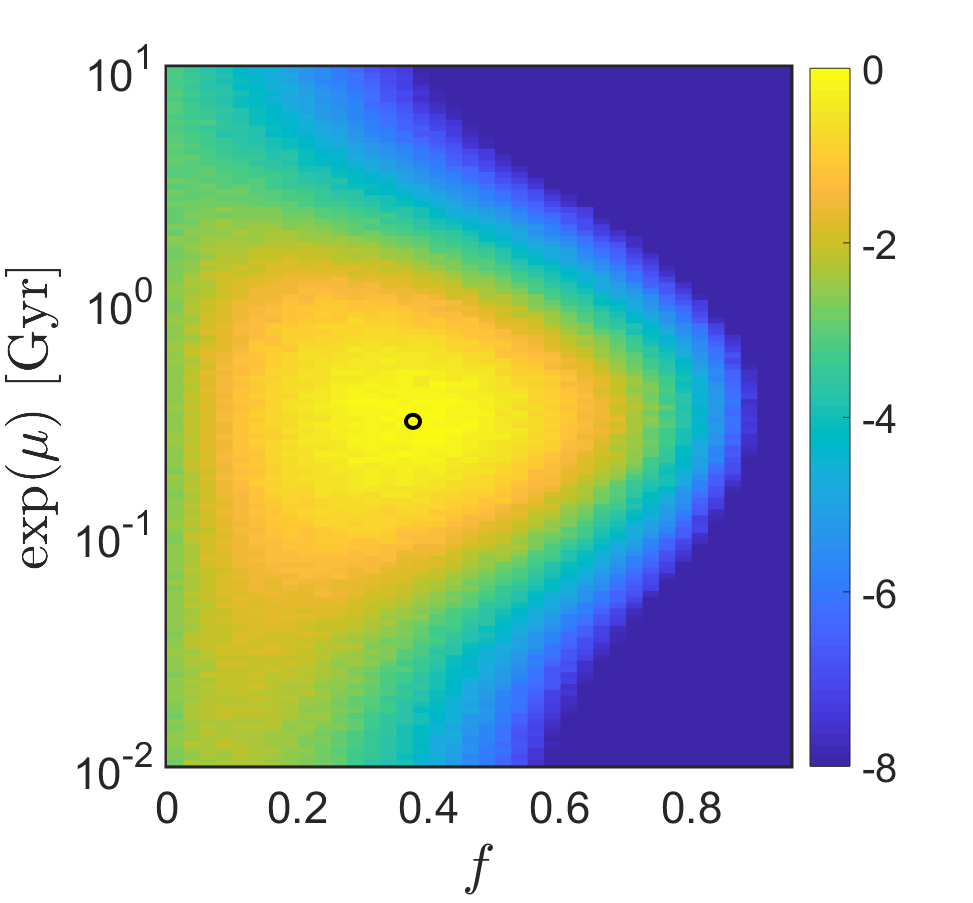}
	\caption{Log-likelihood function for a delay time distribution given by $D_1(t)$ (see \S \ref{sec:shortcomp}). The likelihood is shown as a function of $f,\exp(\mu)$. The maximum is depicted by a black circle for clarity.}
	\label{fig:likelihood}
\end{figure}

The likelihood function for $D_1(t)$ is maximized at $\exp(\mu)=0.3$\,Gyr and $f=0.4$. It is depicted in figure \ref{fig:likelihood}. The resulting DTD is now statistically consistent with the data (see also figure \ref{fig:modelcompare}). The model is insensitive to the exact value of $t_{\rm min}$. Values in the range $0.001\mbox{ Gyr}\leq t_{\rm min} \leq 0.1 \mbox{ Gyr}$ are all consistent with observations. 

Given the  small number of observed systems, the functional form of the DTD cannot be uniquely determined by the available data. As an illustration, we consider a broken power-law:
\begin{eqnarray}
D_2(t)=C	\left\{ \begin{array}{ll}(t/t_0)^{-a} & \mbox{ for  }t_{\rm min}<t< t_0\ , \\
(t/t_0)^{-1}  & \! \! \mbox{ for  }t_0 < t < t_{\rm max}\ .
\end{array} \right. 
\end{eqnarray}
We take, as before, $t_{\rm min}=0.035$ Gyr, $t_{\rm max}=10^6$ Gyr. The free parameters of $D_2(t)$  are $a,t_0$. This model is also consistent. The parameters  $t_0=15\mbox{ Gyr}, a=1.3, t_{\rm min}=0.035$ Gyr  provide a statistically consistent fit to the data. Allowing for varying values of $t_{\rm min}$ we find that consistent solutions can be found as long as $0.02\mbox{ Gyr}\leq t_{\rm min} \leq 0.15\mbox{ Gyr}$. 

\subsection{Comparison between the models}
\label{sec:comp}
To compare $D_1(t),D_2(t),D_3(t)$ to $D\propto t^{-1}$ we perform a likelihood ratio ($LR$) test. The $LR$ is defined as $LR=2\log(L_{\rm max,a}/L_{\rm max,0})$ where $L_{\rm max,a}, L_{\rm max,0}$ are respectively the maximum likelihoods of some model $a$ and of the null hypothesis (model 0). $LR$ follows a $\chi$-squared distribution where the number of degrees of freedom is given by the difference in free parameters between the two models (2 between $D_1,D_2$ and $D\propto t^{-1}$ and 1 between $D_3$ and $D\propto t^{-1}$). For all three models the improvement as compared with a $t^{-1}$ DTD is statistically significant (with a $p$-value for chance coincidence of $p<0.05$). 
Inspection by eye of figure \ref{fig:modelcompare} suggests that the model with an extra short-time  log-normal population looks best. However, this is not statistically significant and it depends on the binning of the data. A  shared property of  all these consistent models is that they all result in a large fraction, $Z_{<1 \rm Gyr}\gtrsim 0.4$,  of BNS with merger times of less than 1\,Gyr.

When a significantly larger number of BNS systems are observed in the future it will be possible to differentiate between more subtle details of the delay time distribution.
In particular this will enable to differentiate between the different models described above.  
A comparison of non-nested models with each other (i.e. comparing any two of $D_1,D_2,D_3$) is impossible with the regular $LR$ test and usually requires a larger data-set to be  meaningful \citep{Vuong89}. 
With coming detections of gravitational waves from BNS mergers, the SFR of the host galaxies will enable us to obtain independent evidence for this fast merging population. This will also enable us to distinguish between the different models described above.  

\subsection{The pulsar's lifetime }
\label{sec:lifetimeLike}
\begin{figure}
	%\centering
	\includegraphics[width=0.24\textwidth]{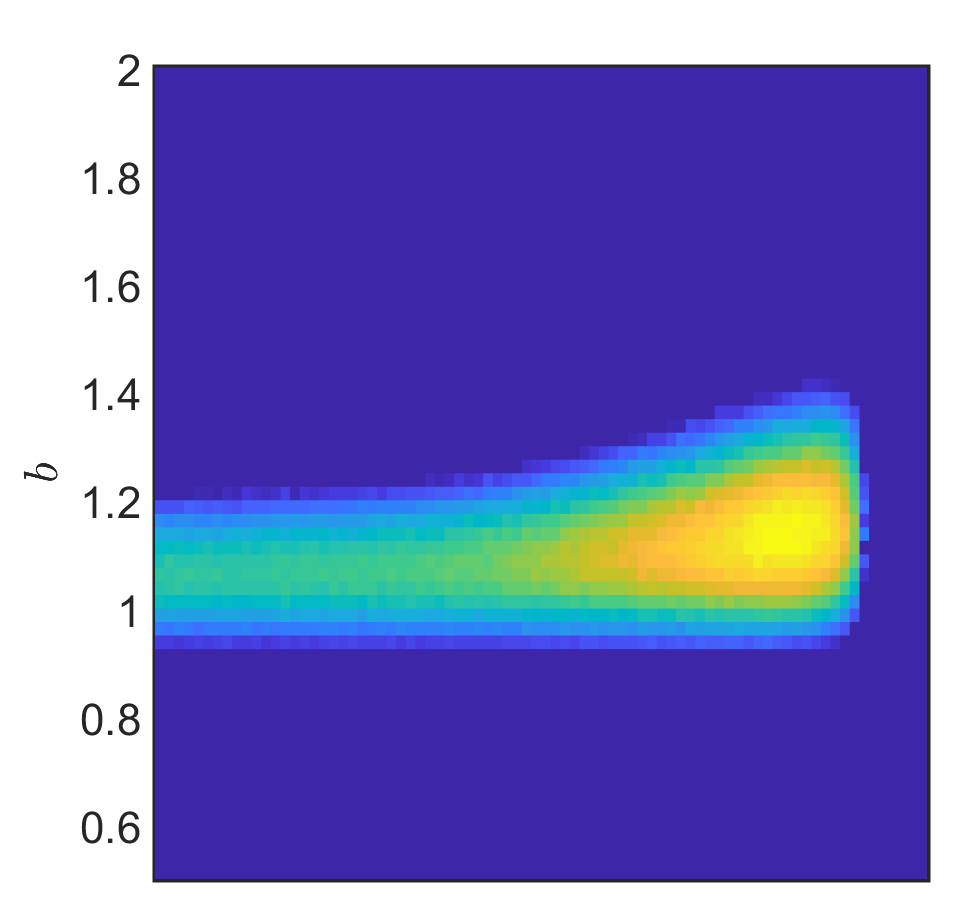}\\
	\includegraphics[width=0.24\textwidth]{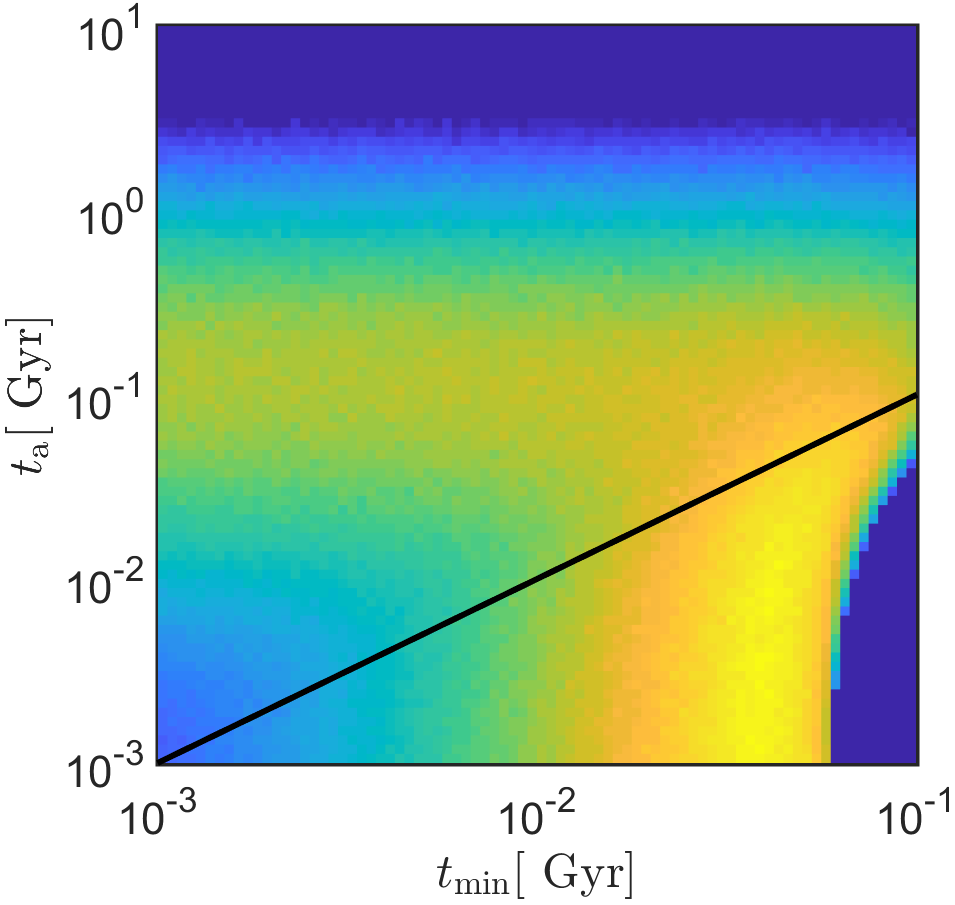}
	\includegraphics[width=0.24\textwidth]{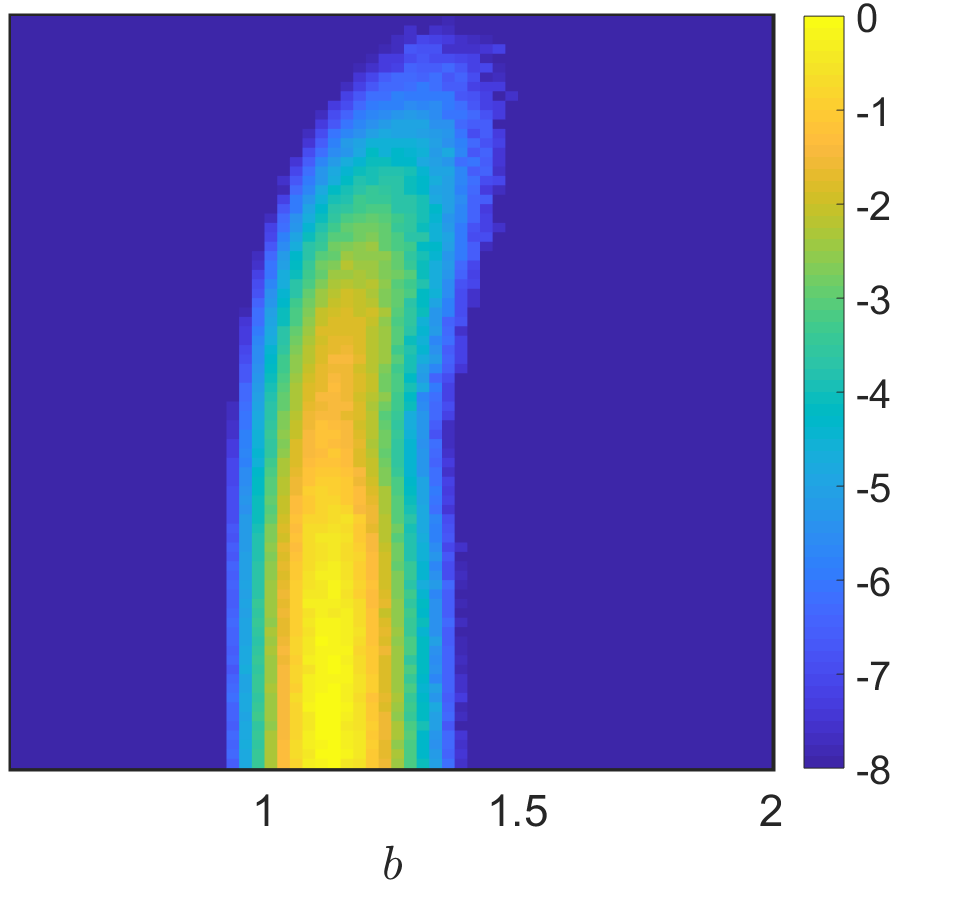}
	\caption{Projections of the 3D Log-likelihood plots in the case of a power-law DTD, $D_3$, where the free parameters are $t_{\rm a},t_{\rm min},b$ and $t_{\rm a}$ the maximal age of detectable pulsars. The solid line in the bottom left figure marks  $t_{\rm a} = t_{\rm min}$. One can see that  $t_{\rm a}< t_{\rm min}$ is preferred by the data even without taking into account the short spin-down times of the observed pulsars. The likelihood is maximized for $b=1.1, t_{\rm min}=0.035\mbox{ Gyr}, t_{\rm a}=0.005\mbox{ Gyr}$.}
	\label{fig:talike}
\end{figure}

Before concluding this section we return once more to the question of the pulsar's lifetime.  To address it we have carried out a likelihood analysis including $t_{\rm a}$ as a free parameter. Figure \ref{fig:talike} depicts the likelihood projections for model $D_3$ but now with $t_{\rm a}$ as a free parameter. Note that in this analysis we did not impose the conditions $t_{\rm age} < P/2\dot P $, as with the very short $P/2\dot P $ of J1906+0746  this would have immediately imposed a very small $t_{\rm a}$ (see \ref{sec:Ages}). Still even without this constraint we find (see figure \ref{fig:talike}) that $t_{\rm a}< t_{\rm min}$ is preferred, where the best fit value is  $\lesssim 0.02$\,Gyr and where in any case the region $t_{\rm a} > 0.1$\,Gyr is practically excluded. 

\section{Total number of surviving and merged Galactic BNS}
\label{sec:numbers}
We turn now to explore the unobserved populations: BNS systems that cannot be observed because their pulsars have died or are not pointing towards us and BNS systems that have already merged.
The fraction of merged systems depends on the formation rate history of BNSs in the MW and on the DTD:
\begin{equation}
\Psi_{\rm merg}= \frac{1}{\int_0^{t_{\rm H}}\!{r(t)} dt} \int_0^{t_{\rm H}}\!{r(t)} P(t_d<t)dt \ ,
\end{equation}
where $P(t_d<t)=\int_0^tD(t)dt$ is the probability of a system born $t$ years ago to have merged by today. A specific case of interest is that of a constant rate $r(t)=\int_0^{t_{\rm H}}\!{r(t)} dt/t_{\rm H}\equiv r_0$. 
In this case, for a power-law distribution, we have: 
\begin{eqnarray}
\Psi_{\rm merg}\!&=&\! \frac{{t_{\rm H}^{2-b}+(1-b)t_{\rm min}^{2-b}} - {(2-b)}t_{\rm H} t_{\rm min}^{1-b}}{(2-b) t_{\rm H}(t_{\rm max}^{1-b}-t_{\rm min}^{1-b})}\nonumber \\ &\approx& 1-\bigg(\frac{t_{\rm min}}{t_{\rm H}}\bigg)^{b-1}\approx 0.5\ , 
\end{eqnarray}
where in the last expression we have used the results for $D_3(t)$ from \S \ref{sec:like}.
We find a similar fraction of merged systems, $\Psi_{\rm merg}(D_1)\approx \Psi_{\rm merg}(D_2)\approx 0.55$,  for the distributions, $D_1(t),D_2(t)$.
We conclude that about half of the BNS systems formed in the history of the Galaxy have already merged.
Using a BNS formation rate within the MW of $\sim 4\times 10^{4}\mbox{ Gyr}^{-1}$ \citep{Kochanek1993,Burgay2003,Kalogera2004,Kim2015,WP2015, Belczynski2018}, this translates to a total of $\sim 2.7\times 10^5$ un-merged BNS systems in the MW.

We next turn to estimate the total number of BNSs with active pulsars that reside in the Galaxy today. As before we take the formation rate to be constant up to $t_{\rm a}$ (see \S \ref{sec:direct}).  The total number of active systems in the MW is then estimated as
\begin{eqnarray}
& N_{\rm active}=\int_0^{t_{\rm a}} r(t)P(t_d>t)dt\approx\\ & r  t_{\rm a}
\left\{ \begin{array}{ll}1 & \! \! t_{\rm a}< t_{\rm min}\\
(t_{\rm a}/t_{\rm min})^{1-b}/(2-b) \! \! & \! \! t_{\rm a}> t_{\rm min}.
\end{array} \right.  \nonumber
\end{eqnarray}
The observable number of systems in the MW is smaller than the number of active ones due to two factors. The first is $\Phi_{\rm beam}$. This is the correction that arises since pulsars are detected by their beamed radio emission. For each pulsar with a period $P$ and temporal beam width $W$, there are approximately $P/2W$ actual pulsars in the Galaxy of the same type. 
Using $W_{50}$ as the beam width (where $W_{50}$ is the width of the pulsar's pulses at a $50\%$ level), we find that for our sample $\Phi_{\rm b}\approx \langle P/(2W)\rangle ^{-1}=0.1$. The other reduction is due to the incompleteness of the BNS pulsar population, $\Phi_{\rm inc}$. For example, \cite{Keane2015} have predicted that SKA could increase the number of known radio pulsars by a factor of $\sim 5-10$, implying $\Phi_{\rm inc}<0.2$. Overall, the observed current number of BNSs in the Galaxy, $N_{\rm obs}\approx 20$ implies $N_{\rm active}=N_{\rm obs} \Phi_{\rm b}^{-1}\Phi_{\rm inc}^{-1}\gtrsim 10^3$. These estimates are consistent with estimates by \cite{Kalogera2003} that calculated there should be $\approx 300$ Galactic systems similar to each B1534+12 and B1913+16.
We can now estimate $t_{\rm a}$,
\begin{eqnarray}
t_{\rm a}\approx\frac{N_{\rm obs}}{ \Phi_{\rm b}\Phi_{\rm inc}r}=0.015 \mbox{ Gyr} \bigg(\frac{0.1}{\Phi_{\rm b}}\bigg)\bigg(\frac{0.03}{\Phi_{\rm inc}}\bigg) \bigg(\frac{4\!\times\! 10^4}{r \mbox{ Gyr}}\bigg) 
\end{eqnarray}
This is consistent with the condition $t_{\rm a}\lesssim t_{\rm min}$ suggested by the distribution of spin-down ages (see \S \ref{sec:Ages}).

\section{Comparison with other results} \label{sec:other}
\begin{figure}
	\centering
	\includegraphics[width=0.39\textwidth]{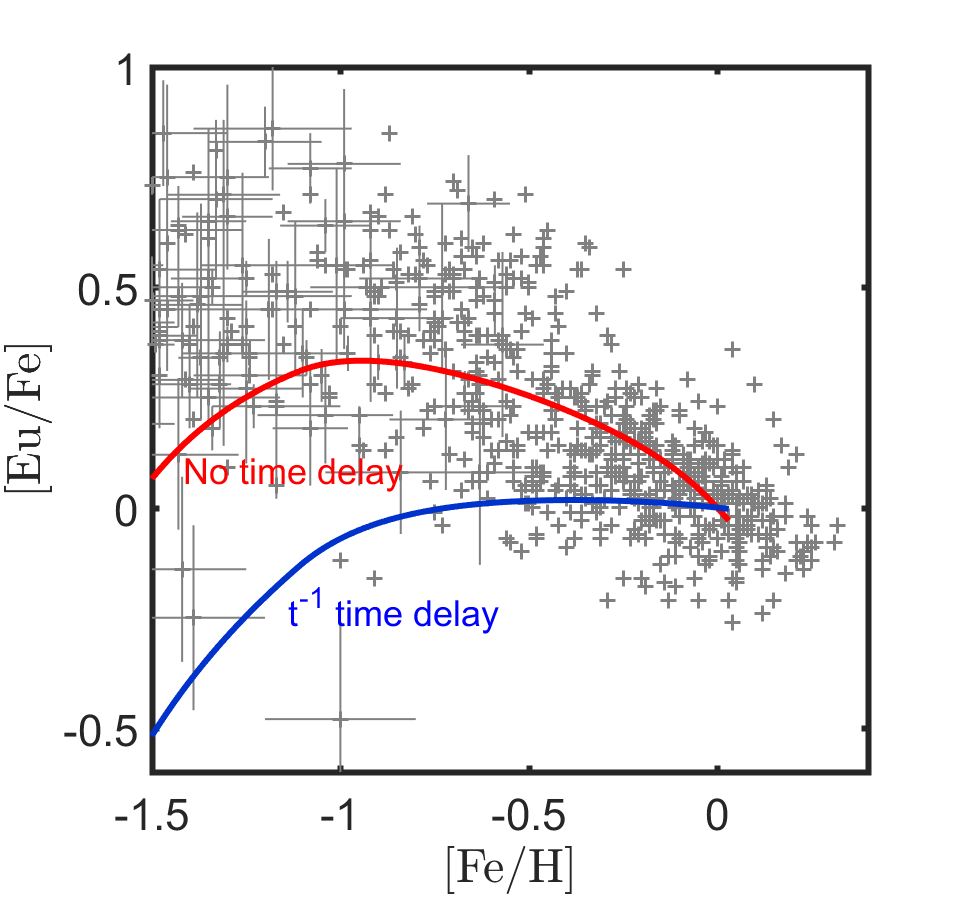}
	\caption{Crosses: abundances of Milky Way stars from the SAGA sample (Suda et al. 2008). Solid lines depict the chemical abundance evolution resulting from the one zone model presented in Hotokezaka, Beniamini $\&$ Piran 2018. The red curve was calculated assuming no delay between formation and merger of BNSs, and the blue with $D\propto t^{-1}$. Clearly short delays improve the agreement with observations.}
	\label{fig:abundanceevolve}
\end{figure}

Our main finding is an excess of rapid mergers over the expected $t^{-1}$ delay time distribution. It turns out that there are other venues that lead to this conclusion. An excess of rapid mergers has also been suggested on completely separate grounds following from observations of $r$-process element abundances.
Specifically, the high abundance of $r$-process elements in the ultra-faint dwarf (UFD) galaxy Ret-II as compared to the upper limits on those abundances in other UFDs suggests a neutron star merger origin \citep{BHP2016}. At the same time, UFDs are known to have ceased their star formation within the first Gyr of their formation \citep{Brown2014,Weisz2015} requiring a rapid merger to enrich the gas before it can be reprocessed into the stars seen in those galaxies today \citep{BHP2016b,Safarzadeh2019}. More generally, if UFD-like galaxies form the main building blocks from which the MW halo population is composed, then rapid mergers are required to explain the high values of $r$-process abundance observed in extremely metal poor stars \citep{Argast2004,Tsujimoto2014,Wehmeyer2015,BHP2016b,Beniamini2018}. Furthermore, a significant population of rapid mergers, coupled with a declining star formation rate over the last few Gyrs implies a declining rate of mergers over the evolution of the MW. This is consistent with the observed rate of deposition of radioactive $^{244}$Pu on Earth today which is considerably lower than its value 4.6 Gyr ago when the solar system was formed \citep{Hotokezaka2015,Wallner2015}. Finally, a population of rapid mergers was shown by \cite{HBP2018} to provide an explanation for the declining rate of [Eu/Fe] as a function of [Fe/H] observed in the MW for $\mbox{[Fe/H]}\gtrsim -1$ which is not trivially explained with a $D\propto t^{-1}$ DTD (see e.g. \citealt{Matteucci2014,Wehmeyer2015,Cote2016,Komiya2016,Simonetti2019}). This result is depicted in figure \ref{fig:abundanceevolve} in which we repeat the one zone chemical evolution model presented in \S 4 of \cite{HBP2018} with the same model parameters (and for the case of a constant Galactic SFR) and changing only the DTD of the NS mergers. With no delays the results track closely the observed abundance patterns in Milky Way stars. However, a $t^{-1}$ DTD, clearly evolves too shallowly.

As mentioned in \S \ref{sec:Intro}, using the sGRB redshift and peak flux distribution, \cite{WP2015} have found that $dN/dt_{\rm m}\propto t_{\rm m}^{-0.8 {\pm 0.25}}$. There is some tension between this result and ours. However, \cite{WP2015} allow in their analysis for the existence of a second population that follows the SFR. They find that this population includes about $1/3$ of sGRBs and interpret this as these bursts being ``short duration" Collapsars \citep{Bromberg2013}. These findings are consistent with our results if a significant fraction of these are genuine sGRBs. 

We can compare our results to the implications of the first NS merger observed in gravitational waves, GW170817. As mentioned earlier the time delay for GW170817 is estimated at $t_{\rm 170817} \approx 1-10$ Gyr. Thus, GW170817, can be used to put an {\it upper} limit on the rate of short mergers due to the following argument. The inferred rate of BNS mergers and the amount of $r$-process mass estimated from this single event are already large enough to dominate the production of $r$-process elements in the Universe (see e.g. \citealt{HBP2018} for a detailed discussion).
In fact, they are already somewhat larger than those inferred from the total $r$-process mass in the MW (albeit with still relatively large errors).
If the delay time distribution is significantly steeper than $t^{-1}$, then the inferred rate of overall mergers (and in turn their contribution to the total $r$-process mass) would have been much larger in the past (  $\propto (t_{\rm 170817}/t_{\rm min})^{b-1}$).  Applying a conservative limit, that the overall merger rate should not be more than 5 times larger than that inferred by GW170817, we find
$b<1+{\log(5)}/{\log(t_{\rm 170817}/t_{\rm min})} $
This leads to $b<1.5$ for $t_{\rm 170817}=1$\,Gyr and $b<1.3$ for $t_{\rm 170817}=10$\,Gyr. These limits are consistent with the level of enhancement of short mergers found in the present work. 

\section{Discussion}
\label{sec:discussion}
We have explored the distribution of delay times (DTD) between formation and merger of binary neutron star (BNS) systems.
We have shown that the times since formation of the observed population of BNS systems, that are dictated by the lifetime of the pulsars, are sufficiently short so that the ages of the observed BNS systems have little to no effect on the observed merger time distribution. This distribution then directly traces the intrinsic DTD distribution. Using a maximal likelihood we found several distributions that fit the observed data.

At long times ($t\gtrsim 1$\,Gyr), the DTD is consistent with the theoretically expected $D\propto t^{-1}$. At shorter times, however, there is a statistically significant excess of rapidly merging systems (almost half of the systems merge in less than 1 Gyr). We have shown that the finite pulsar's lifetime cannot explain this excess which must be intrinsic. We explored a few models with this characteristic.
One such model that provides a good description of the data is a DTD where  40\% of the systems arise from a log-normal distribution with $\mu=\log(0.3/\mbox{ Gyr}), \sigma=1$ while the other  60\%  arise from a $t^{-1}$ distribution, with minimum and maximum merger times of $t_{\rm min}=0.035 \mbox{ Gyr}, t_{\rm max}=10^6\mbox{ Gyr}$. However there is not enough data to distinguish between this distribution and distributions that have other functional forms that include an excess of short merger time systems. 

The observed chemical evolution of heavy $r$-process elements in the Milky Way is estimated from the ratio of [Eu/Fe] as compared with  [Fe/H]. The decrease in [Eu/Fe] for $\mbox{[Fe/H]}\gtrsim -1$ has been puzzling within the context of BNS as sites of $r$-process nucleosynthesis as it is not trivially explained with a $D\propto t^{-1}$ DTD (see e.g. \citealt{Matteucci2014,Wehmeyer2015,Cote2016,Komiya2016,Simonetti2019}). As mentioned earlier this  population of rapid mergers    provide a natural  explanation for this puzzle  \citep{HBP2018}.

The range of merger times, from $t_{\rm min} \!\approx\! 0.035~ (0.001-0.15)$\, Gyr\footnote{The values in bracket correspond to upper and lower estimates of $t_{\rm min}$}  to $t_{\rm max}\!\approx \!10^6$\, Gyr corresponds to initial orbital separations (for a canonical $1.4-1.4 M_\odot$ binary) of   $7 ~(3-10)\cdot 10^{10} $\,  cm to $ 5 \cdot 10^{12}$\, cm. The minimal value is the most interesting as it provides an indication on the state of the system just before it collapses to form a neutron star. This distance is about  a solar radius. Since the progenitor is essentially heavier than that, it implies that the progenitor must have lost a significant fraction of its envelope just before the collapse, or that its companion was within the envelope during the collapse.

Our inferred DTD suggests that roughly half of the BNS systems ever formed  have merged by today. This could lead to $\approx 3\times 10^5$ unmerged BNS systems in the MW. That being said, the vast majority of those systems are not expected to host active pulsars.

The departure of the DTD from a simple $t^{-1}$ may point towards different channels of BNS formation.
Indeed \cite{BP2016} have shown that there are two classes of collapses that lead to the formation of the second NS in a BNS system. In roughly $60-70\%$ of the systems, the collapse involves only a small amount of mass ejection ($\Delta M\lesssim 0.5 M_{\odot}$) and a weak kick ($v_k\lesssim 30 \mbox{km/sec}$), while in the rest of the systems the collapse involve larger ejected masses and kicks. The first type of collapse results in low eccentricity ($e\approx 0.1$) systems, while the second can result in much more eccentric orbits. Since there appears to be no correlation between $e$ and $t_{\rm m}$ (see table \ref{tbl:sample}), it seems unlikely that the excess of rapidly merging systems is related to the nature of the kicks. An alternative possibility is that the excess is related to the initial separation at birth (see also \citealt{Belczynski2006}), which in turn may indicate a different stellar evolutionary part  involving a larger shrinkage of the orbit in a common envelope phase.

GW detectors are expected to detect many more BNS mergers in the coming years \citep{GW170817}. Although the coincident detection of GRBs are predicted to be much less common \citep{Beniamini2019}, we may expect the detection of the isotropic Macronovae in most cases. These will provide us with further data concerning the rate and $r$-process yield of BNS mergers. The SFR history of the host galaxies could reveal information on the DTD. We expect that at least 50\% of the hosts will show evidence of recent SFR. 
Advanced radio telescopes such as SKA would increase the sample of Galactic BNSs and this will enable us to improve the statistics explored here.

\section*{Acknowledgments}
We thank Ehud Nakar,  Bernard Schutz and Ben Shenhar for helpful discussions. TP was partially supported by an ERC advanced grant TReX and by the CHE-ISF center for excellence in Astrophysics.

	%	\bibliographystyle{mnras}
	
	%%%%%%%%%%%%%%%%% APPENDICES %%%%%%%%%%%%%%%%%%%%%
	% Don't change these lines
	\bsp	% typesetting comment
	\label{lastpage}
\end{document}